\documentclass[twocolumn,amssymb, aps, prc, longbibliography]{revtex4}

\usepackage{graphicx}
\usepackage{amsmath}
\usepackage{multirow}
\usepackage{xcolor}

\begin{document}

\begin{abstract}
Particle-decaying states of the light nuclei $^{11,12}$N and $^{12}$O were studied using the invariant-mass method. The decay energies and intrinsic widths of a number of  states  were measured, and the momentum correlations of three-body decaying states were considered. A second 2$p$-decaying $2^{+}$ state of $^{12}$O was observed for the first time, and a higher energy $^{12}$O state was observed in the 4$p$+2$\alpha$ decay channel. This 4$p$+2$\alpha$ channel also contains contributions from fission-like decay paths, including $^{6}$Be$_{g.s.}$+$^{6}$Be$_{g.s.}$. Analogs to these states in $^{12}$O were found in $^{12}$N in the 2$p$+$^{10}$B and 2$p$+$\alpha$+$^{6}$Li channels. The momentum correlations for the prompt 2$p$ decay of $^{12}$O$_{g.s.}$ were found to be nearly identical to those of $^{16}$Ne$_{g.s.}$, and the correlations for the new 2$^{+}$ state were found to be consistent with sequential decay through excited states in $^{11}$N. The momentum correlations for the 2$^+_1$ state in $^{12}$O provide a new value for the $^{11}$N ground-state energy. The states in $^{12}$N/$^{12}$O that belong to the $A$=12 isobaric sextet do not deviate from the quadratic isobaric multiplet mass equation (IMME) form.
\end{abstract}

\title{Particle decays of levels in $^{11,12}$N and $^{12}$O investigated with the invariant-mass method } %

\author{T.~B.~Webb}%
\email[Corresponding author: ]{tbwebb@go.wustl.edu}
\affiliation{Departments of Chemistry and Physics, Washington University, St. Louis, MO 63130}
\author{K.~W.~Brown}
\affiliation{National Superconducting Cyclotron Laboratory, Michigan State University, East Lansing, MI 48824.0}
\author{R.J.~Charity}
\author{J.M.~Elson}
\author{D.~E.~M.~Hoff}
\author{C.~D.~Pruitt}
\author{L.~G.~Sobotka}
\affiliation{Departments of Chemistry and Physics, Washington University, St. Louis, MO 63130}
\author{J.~Barney}
\author{G.~Cerizza}
\author{J.~Estee}
\author{G.~Jhang}
\author{W.~G.~Lynch}
\author{J.~Manfredi}
\author{P.~Morfouace}
\author{C.~Santamaria}
\author{S.~Sweany}
\author{M.~B.~Tsang}
\author{T.~Tsang}
\author{S.M.  Wang}
\author{Y.~Zhang}
\author{K.~Zhu}
\affiliation{National Superconducting Cyclotron Laboratory, Michigan State University, East Lansing, MI 48824.0}
\author{S.~A.~Kuvin}
\author{D.~McNeel}
\author{J.~Smith}
\author{A.~H.~Wuosmaa}
\affiliation{Department of Physics, University of Connecticut, Storrs, CT 06269}
\author{Z.~Chajecki}
\affiliation{Department of Physics, Western Michigan University, Kalamazoo, MI 49008}
\date{present}%
\maketitle

\section{INTRODUCTION}
Isotopes near and beyond the drip lines can show interesting properties such as spatially extended halos, prompt two-nucleon emission, and low-energy intruder states.  They also present a theoretical challenge as coupling to the  continuum needs to be included which can result in clusterization \cite{Okolowicz:2012} and the breaking of isospin symmetry due to Thomas-Ehrman effects \cite{Thomas:1951,Ehrman:1951,Grigorenko:2002,Michel:2010}. 

Experimentally, the invariant-mass method can play an important role in the study of such effects, finding new levels, and characterizing their decays. It is particularly useful in 3-body decays as the momentum correlations between the decay products can capture information about the decay dynamics and the structure of the parent states. These correlations can be characterized by a two-dimensional plot of kinematic relations between the decay fragments.  High-statistics measurements of such plots for the democratic 2$p$ decay of the $^6$Be  and $^{16}$Ne ground states  have been made and successfully compared to three-body calculations \cite{Egorova:2012,Brown:2015}. For the intermediate case, $^{12}$O$_{g.s.}$, only a 1-dimensional distribution has been measured with lower statistics \cite{Kryger:1995}. 

The structure of $^{11}$N is important for excited $^{12}$O levels that sequentially 2$p$ decay. Knowledge of the energies, widths, and decay paths of these states will lead to better understanding of this sequential 2$p$ decay and constrain theoretical calculations.

The isobaric analog of $^{12}$O in $^{12}$N was also shown to undergo prompt 2$p$ decay to the isobaric analog state (IAS) in $^{10}$B \cite{Jager:2012}. This represented a second case of prompt 2$p$ decay from one IAS to another IAS, with  the first case being $^{8}$B$_{IAS}\rightarrow$2$p$+$^{6}$Li$_{IAS}$ \cite{charity_carbon2011,Brown:2014}. It is interesting to determine if these analog 2$p$ decays have the same momentum correlations as their ground-state cousins. $^8$B$_{IAS}$ is the analog of $^8$C$_{g.s.}$, which decays by two sets of prompt 2$p$ decays \cite{charity_carbon2011}. It is difficult to separate the protons from the two steps, making comparison of the analog 2$p$ decays difficult. On the other hand, $^{12}$O$_{g.s.}$ and its analog $^{12}$N$_{IAS}$  present no such difficulties, as there is only one step of 2$p$ decay in both systems. It is also interesting to see if higher-lying analog states undergo prompt or sequential 2$p$ decay to the $^{10}$B$_{IAS}$.      
 
This work is part of a larger joint experimental and theoretical study of $^{12}$O and its neighbors with the aim of addressing these issues. The present paper mostly presents experimental results with comparisons to other experimental values and some previous theoretical predictions. Comparisons with predictions of the Gamow coupled-channel (GCC) method \cite{Wang:2017,Wang:2018} are given in Refs.~\cite{Webb:2019,Wang:2019} and this work, and further comparisons will be elaborated on in future papers when the model is extended to include the dynamics of the 2$p$ decay.

\section{EXPERIMENTAL METHODS}
The data presented were obtained at the National Superconducting Cyclotron Laboratory at Michigan State University, which provided a ${}^{16}$O primary beam at 150 MeV/A. The primary beam bombarded a 193 mg/cm$^{2}$ ${}^{9}$Be target, and ${}^{13}$O fragments were selected by the A1900 magnetic separator. Upon extraction from the separator, this 69.5 MeV/A secondary beam had a purity of only 10\%. To remove the substantial contamination, the beam was sent into an electromagnetic time-of-flight filter, the Radio Frequency Fragment Separator \cite{bazin2009}, and emerged with a purity of 80\%. The final secondary beam impinged on a 1-mm-thick ${}^{9}$Be target  with a rate of $10^3$ pps and the charged particles produced were detected in the High Resolution Array (HiRA) \cite{WALLACE2007302} consisting of 14 $\Delta$\textit{E-E} [Si-CsI(Tl)] telescopes 85 cm downstream of the target. The array subtended a polar angular range of 2.1$^{\circ}$ to 12.4$^{\circ}$ and a solid angle of 78 msr. Each telescope consisted of a 1.5-mm-thick, double-sided Si strip $\Delta E$ detector followed by a 4-cm-thick, CsI(Tl) $E$ detector. The $\Delta E$ detectors are 6.4~cm $\times$ 6.4~cm in area, with each of the faces divided into 32 strips. Each $E$ detector consisted of four separate CsI(Tl) elements each spanning a quadrant of the preceding Si detector. Signals produced in the 896 Si strips were processed with the HINP16C chip electronics \cite{Engel07}.

The energy calibration of the Si detectors was obtained with a $^{232}$U 
$\alpha$-particle source. The particle-dependent energy calibrations of the CsI(Tl)
detectors were achieved with cocktail beams selected with the A1900 separator.
These included protons (80~MeV), $N$=$Z$ fragments ($E/A$=80~MeV), and $^9$C ($E/A$=82.9~MeV). Lower-energy points were obtained using 2.2-, 4.9-, and 9.6-mm-thick Al degraders. Calibrations for $^{10}$C and  $^{9}$Be fragments were obtained 
from interpolating and extrapolating these results.

\begin{figure}
\includegraphics[scale=0.3]{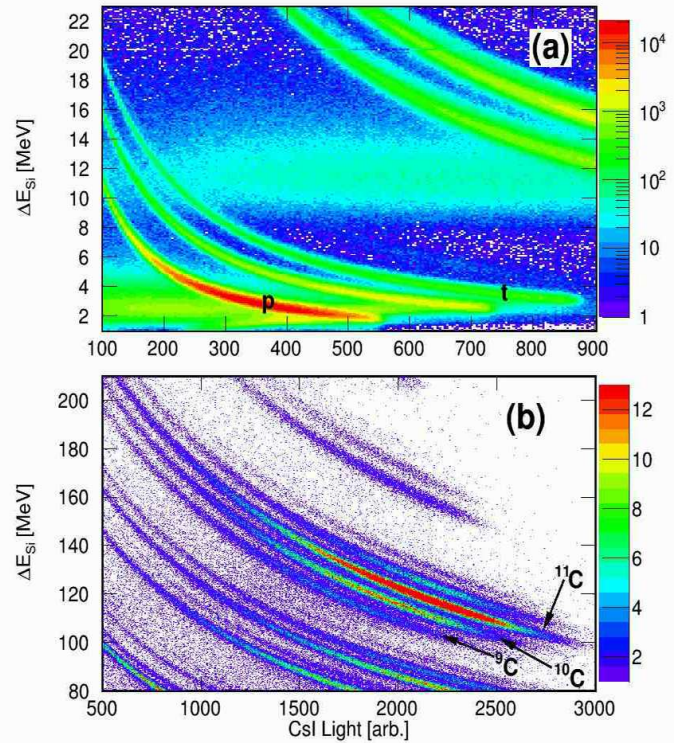}
\caption{Zoomed-in views of a typical $E-\Delta E$ plot 
from one of the telescopes showing particle identification. (a) shows the region of H isotopoes with bands corresponding to $p$, $d$, and $t$ fragments clearly separated. The $z$ axis is shown in a logarithmic scale so that all three isotopes are clearly visible. (b) shows the region of C fragments with bands associated with isotopes from $^9$C to $^{13}$C visible. The $z$ axis in this case is shown with a linear scale.}  
\label{fig:dEE}
\end{figure}

Particle identification was obtained from  $E-\Delta E$ plots where the $y$-axis gives the energy lost in the silicon detector  while the $x$-axis is related to the residual energy which is deposited in the CsI(Tl) scintillators ($E$).     Figure~\ref{fig:dEE} shows zoomed-in regions of such a plot for a typical telescope  centered on the regions containing the H [Fig.~\ref{fig:dEE}(a)] and C  [Fig.~\ref{fig:dEE}(b)] isotopes. Isotopes separation is obtained in all cases.

For the normalization of cross sections, the beam particles were counted using a thin plastic-scintillator foil at the  exit of the  Radio Frequency Fragment Separator. The purity of the beam and the beam flux loss due to its transport to the target were determined periodically in the experiment by temporarily placing a CsI(Tl) detector in the target position. These fluxes were also corrected for the detector dead time measured with a random pulse generator.

\begin{figure}
\includegraphics[scale=0.4]{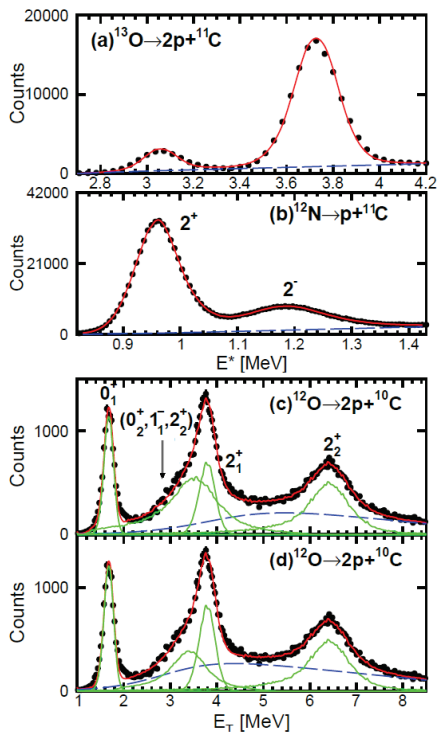}
\caption{Excitation-energy spectra of calibration peaks for (a) the first 2$p$-decaying levels of ${}^{13}$O and (b) the first two excited states of ${}^{12}$N reconstructed from transversely decaying 2\textit{p}+${}^{11}$C and \textit{p}+${}^{11}$C events respectively. The solid red curves show the results of Monte Carlo simulations incorporating the detector response and known level properties. The blue dashed curves are fitted estimations of the background. Total decay energy spectra (c), (d) of ${}^{12}$O reconstructed from transversely decaying 2\textit{p}+${}^{10}$C events. (c) shows a fit at the upper limit of the uncertainty for the 0$^{+}_{2}$ width while (d) shows the fit at the lower limit. The red solid curve shows the summed results of Monte-Carlo simulations of each resonance, shown by the solid green curves. The blue dashed curve shows the fitted background. A $\left|\cos\theta_{C}\right|< $0.2 gate was applied for all spectra.}
\label{fig:calibrationsandO12}
\end{figure}

\section{INVARIANT-MASS METHOD}
In the decay of a parent state into two or more pieces,
the  mass of the parent can be obtained from the invariant mass of the decay products. Experimentally we measure the invariant mass $M_{inv}$ of the charged particles detected in the HiRA array. For example 2$p$+$^{10}$C events associated with the decay of $^{12}$O levels are obtained from two protons and a $^{10}$C fragment detected in coincidence where each fragment's $Z$ and $A$ are identified from the $E-\Delta E$ plots. The latter requires that each fragment be detected in a separate CsI(Tl) crystal and four-fold or higher-order coincidences containing 2$p$+$^{10}$C subevents are rejected.
For $^{11}$N resonances decaying to the $p$+$^{10}$C channel we can similarly look at double coincidences of a proton and a $^{10}$C fragment. However some of these events are in  reality  2$p$+$^{10}$C decays of $^{12}$O where one of the protons was not detected.  In this case, background from the latter events must be considered (sec.~\ref{sec:N11}). The $p$+$^{10}$C channel is the only channel in this work where a background subtraction was necessary as the corresponding backgrounds from high-order coincidences for other channels was found to be negligible.

Rather than using the $M_{inv}$, it is often more convenient to use the total kinetic energy ($E_{T}$) of the fragments in the center-of-mass frame of these fragments
\begin{equation}
E_{T} = M_{inv} - \sum M_i
\end{equation}
where $M_i$ are the rest masses of the detected fragments. If the fragments are from the decay of a resonance, $E_{T}$ represents the kinetic energy released in the decay of that resonance.  Alternatively if the ground-state rest mass $M_{g.s.}^{parent}$ of the parent is well determined we use
\begin{equation}
 E^*_{n\gamma} =M_{inv} - M_{g.s.}^{parent}.
\end{equation}
This quantity only represents the true excitation energy if all of the detected fragments were produced in their ground states rather than particle-stable excited states. For some of the detected fragments considered in this paper ($^6$Li, $^{10}$B, and $^{10}$C), there are $\gamma$-decaying excited states, and thus the true excitation energy is obtained by adding the sum of the $\gamma$-rays energies, i.e. $E^* = E^*_{n\gamma} + \sum_i E_{\gamma}^i$. Both $^6$Li and $^{10}$C have only one $\gamma$-decaying excited state, limiting the number of possible excitation values to two for an event that includes one of these fragments.

The largest degrading factor to the invariant-mass resolution is from the energy loss of the decay fragments as they leave the target material \cite{Charity:2019}. In particular it is the change in the relative velocities of the fragments that is important, with protons having much smaller velocity losses compared to the heavier cores. The difference in the velocity losses depends on the interaction depth in the target material, which is not known. We make an average correction to the velocities by assuming the reaction occurs in the center of the target material, which gives the correct average invariant mass. The contribution of this effect to the resolution can be minimized by selecting events based on the emission angle $\theta_C$ of the core relative to the beam axis in the parent's center-of-mass frame \cite{Charity:2019}. Transverse emissions, where $\theta_C\sim$90$^\circ$ ($\cos\theta_C\sim$0), have the best resolution, and the width of the gate around this value is optimized for each spectrum based on the statistics and the presence of narrow peaks. The gates used are specified in the figure captions.

Peaks found in the experimental invariant-mass spectra were fit to extract the resonance energy and intrinsic width assuming some intrinsic line shape.  The experimental resolution and efficiency is included via Monte-Carlo simulation \cite{Charity:2019}. These simulations are benchmarked by comparison to well-known narrow resonances with high statistics.  

 Figures \ref{fig:calibrationsandO12}(a) and (b) show the excitation-energy spectra for particle-unbound states of $^{13}$O from detected 2$p$+$^{11}$C events and of $^{12}$N from $p$+$^{11}$C events respectively. The curves show fits assuming Breit-Wigner intrinsic line shapes. The fitted parameters for the 2$^+_1$ and 2$^-_1$ states of $^{12}$N are shown in Table~\ref{tbl:calibration} and compared to the ENSDF values. The reported width of the 2$^+_1$ state is much smaller than the experimental resolution, so the width of this peak is essentially only from the experimental resolution. The simulated peak calculated with $\Gamma$=0 reproduces the observed peak correctly.  

The peaks in the 2$p$+$^{11}$C spectrum have only been seen in one previous study  \cite{Sobotka:2013}. Again we obtained excellent reproduction of these  peaks with $\Gamma$=0 as their width is also dominated by the experimental resolution.  The centroid for the lower-energy peak is consistent with the previous value, while the higher-energy is slightly higher (see Table \ref{tbl:calibration}). Based on past experience, we expect a systematic error of around 10~keV \cite{Charity:2019} which is not included in the uncertainties listed in the following tables. Our reproduction of these known resonances assures us that our simulations correctly incorporate the detector response.

\begin{table}
\caption{Comparison of fitted and literature decay energies and widths of calibration resonances used for detector characterization. Literature values for ${}^{12}$N states come from \cite{AJZENBERGSELOVE1990} and values for ${}^{13}$O states from \cite{Sobotka:2013}}
\label{tbl:calibration}
\begin{ruledtabular}
\begin{tabular}{c c c c c c}
Nuclide & $J_{\pi}$ & $E^{*}_{lit}$ [MeV] & $\Gamma_{lit}$ [keV] & $E^{*}$ [MeV] & $\Gamma$ [keV]\\
\hline 
${}^{12}$N & $2^{+}$ & 0.960(12) & \textless20 & 0.962(3) &  \\
${}^{12}$N & $2^{-}$ & 1.179(17) & 55(14) & 1.186(7) & 77(12) \\
${}^{13}$O & & 3.025(16) & \textless0.050 & 3.038(9) &  \\
${}^{13}$O & & 3.669(13) & \textless0.050 & 3.701(10) & \\
\end{tabular}
\end{ruledtabular}
\end{table}
\section{INVARIANT-MASS SPECTRA} \label{sec:invariantM}
Invariant-mass spectra for all the isotopes examined in this work are presented in this section. The level schemes of Fig.~\ref{fig:level_2pB10} contain new and previously known levels, and can be used to aid in understanding the discussions.  
\subsection{$^{11}$N structure} \label{sec:N11}
\begin{figure}
\includegraphics[scale=0.35]{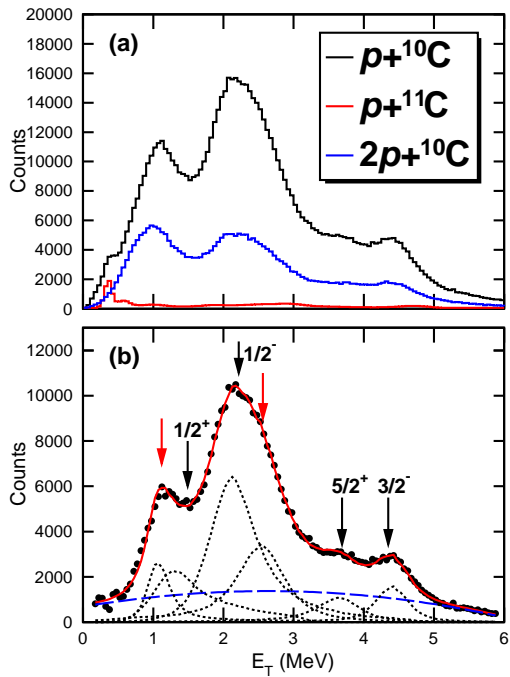}
\caption{$p$+$^{10}$C invariant-mass spectra. (a) The raw spectrum is shown by the black histogram. The red and blue histograms show estimates of the  contamination from $p$+$^{11}$C events where the $^{11}$C fragment was misidentified as a $^{10}$C and from $^{12}$O$\rightarrow$2$p$+$^{10}$C events where one of the protons was not detected. (b) Fit to the contamination-subtracted spectrum with known $^{11}$N peaks in ENSDF indicated by the labeled black arrows and new peaks by the unlabeled red arrows. The fitted peaks for individual levels are shown as the dotted black curves, while the blue dashed curve is the fitted background. A $\left| \cos\theta_{C}\right|<$0.5 gate was applied for all spectra.}
\label{fig:N11}
\end{figure}

The structure of $^{11}$N is important for $^{12}$O decay as its levels are possible intermediate states in sequential 2$p$ decays. The $E_T$ spectrum from detected $p$+$^{10}$C events is shown as the black histogram in Fig.~\ref{fig:N11}(a). From this spectrum we have subtracted two types of contaminations. First, a small fraction of $^{11}$C fragments leak into the $^{10}$C gate. To estimate the spectrum from such events, we have taken the detected $p$+$^{11}$C events and analyzed them as if the $^{11}$C fragment was $^{10}$C. This spectrum is scaled by the probability of misidentification determined from the calibration beams ($\sim$0.5\%) to give the red histogram in Fig~\ref{fig:N11}(a). Its magnitude is only significant at very small values of $E_T$. Secondly, $^{12}$O$\rightarrow$2$p$+$^{10}$C  events in which one of the protons was not detected will also contribute to the detected $p$+$^{10}$C events. To account for this, we have taken the detected 2$p$+$^{10}$C events and randomly thrown away one of the protons, scaling the subsequent $p$+$^{10}$C invariant-mass spectra (blue histogram) based on the results of the Monte-Carlo simulation to appropriately represent the number of contaminated events from this process.

\begin{figure}
\includegraphics[scale=0.4]{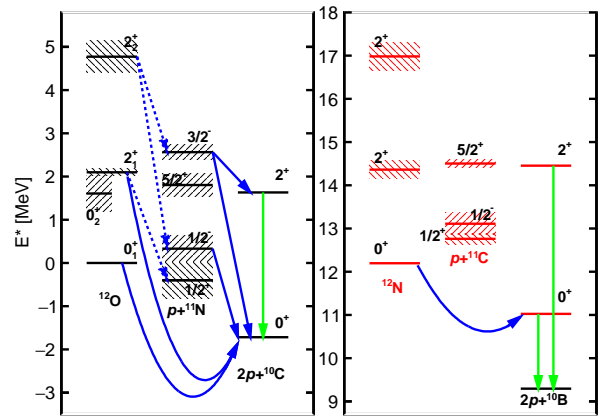}
\caption{Levels of interest in the 1$p$ decay of $^{11}$N states, 2$p$ decay of $^{12}$O states, and their analogs in $^{12}$N. Green arrows show $\gamma$-ray decays of importance, while observed 1$p$ decays are shown as the straight blue arrows. The dashed curves show possible sequential decay paths of $^{12}$O states investigated in this work. The curved blue arrows indicate direct 2$p$ decay. The 0$^+_2$ $^{12}$O state from \cite{Suzuki:2016} is also shown. For the decay of $^{12}$N levels, only isospin-allowed intermediate and final states are shown.} 
\label{fig:level_2pB10}
\end{figure}

The contamination-subtracted spectrum is shown as the data points in Fig.~\ref{fig:N11}(b) and largely represents the yield from the knockout of both a proton and a neutron from the projectile. The $^{12}$O$\rightarrow$2$p$+$^{10}$C events can also give information of $^{11}$N as shown later (Sec.~\ref{sec:corr}).  The labeled black arrows show the locations of levels listed in the ENSDF database. A strong peak for the $J^\pi$=1/2$^{-}$ first-excited state is observed, as is a smaller peak for the $J^\pi$=3/2$^{-}$ state. The  1/2$^{+}$ (ground state) and 5/2$^+$ levels, if present, are not resolved.  The unlabeled red arrows show the locations of new peaks necessary to fit the spectrum. The  dominant 1/2$^-$ peak has a high-energy tail  which cannot be explained by its expected line shape. In fitting the spectrum, we have assumed a new level just above the 1/2$^-$ peak in order to reproduce the observed shape.  More remarkable is the presence of a low-energy peak ($E_T\sim$1.1~MeV) well below the nominal ground-state energy of $E_T$=1.49~MeV \cite{ENSDF}. It is unlikely that this is a true $p$+$^{10}$C$_{g.s.}$ resonance, as we would expect it to have been observed already. Rather the more reasonable explanation is that it represents a $p$ decay to the $J^\pi$=2$^+$ first-excited state of $^{10}$C which gamma decays ($E_{\gamma}$=3.354~MeV, see Fig.~\ref{fig:level_2pB10}). Adding the $\gamma$ energy  moves this peak to the same energy as the 3/2$^-$ level. Thus we conclude the $J^\pi$=3/2$^{-}$ level has proton decay branches to both the ground and first excited state of $^{10}$C.

The solid curve in Fig.~\ref{fig:N11}(b) shows a fit to the invariant-mass spectrum with 5 levels (6 peaks), including the known states indicated by the black arrows and one new level. $R$-matrix line shapes are used for all levels except for the new state where a simple Breit-Wigner shape is assumed, as the decay orbital angular momentum is unknown. The energies and widths of the  unresolved 1/2$^+$ and 5/2$^+$ states were fixed to the ENSDF values, but  their presence allows for a good fit with a smooth background (blue dashed curve). The fitted  branching ratio of the 3/2$^-$ state is $\Gamma_{p_0}/(\Gamma_{p_0}+\Gamma_{p_1}$) = 0.64(6). Fitted energies and widths are listed in Table~\ref{tbl:N11}. The new $E_{T}$=2.563-MeV state, can correspond to either an $E^*$=1.073-MeV or 4.416-MeV level depending on whether the proton decay is to the ground or first excited state of $^{10}$C. No known mirror states in $^{11}$Be exist for the lower possibility, but a couple of candidates exist for the higher value, suggesting that the decay is to the $^{10}$C excited state.  

\begin{table}
\caption{Level parameters obtained from the fit to the $p$+$^{10}$C invariant-mass spectrum of Fig.~\ref{fig:N11}(b).}
\label{tbl:N11}
\begin{ruledtabular}
\begin{tabular}{c c c c c c}
           &       \multicolumn{2}{c}{ENSDF} & \multicolumn{3}{c}{This Work} \\
 $J_{\pi}$ & $E_{T}$ [MeV] & $\Gamma$ [keV] & $E_{T}$ [MeV]  & $\Gamma$ [keV] & $E^*$\footnotemark[3] [MeV] \\

\hline 
1/2$^{+}_{1}$ & 1.49(6)   &  830(30)   & 1.378(5)\footnotemark[1]  & 780(10)\footnotemark[1] & 0.\\
1/2$^{-}_{1}$ & 2.220(30) &  600(100)    & 2.216(5)  & 721(13) & 0.838 \\
   ?     &           &            &  2.563(10) & 697(32) &  1.185 or 4.539 \\
5/2$^+$ & 3690(30)\footnotemark[2]  &  540(40)\footnotemark[2]   & &  & 2.312 \\
3/2$^{-}$ & 4.350(3)  &  340(40) & 4.475(5) & 340\footnotemark[2]  & 3.097\\
\end{tabular}
\end{ruledtabular}

\footnotetext[1]{From an the analysis of $^{12}$O$\rightarrow$2$p$+$^{10}$C events in Sec.~\ref{sec:corr}.}
\footnotetext[2]{Parameters fixed to ENSDF values in fit.}
\footnotetext[3]{Based on our new value for the ground-state energy.}
\end{table}

\subsection{$^{12}$O Structure}
\subsubsection{2$p$+$^{10}$C channel}
\label{sec:O12}
The $E_T$ spectrum for the 2$p$-decay of ${}^{12}$O is shown in both Figs. \ref{fig:calibrationsandO12}(c) and (d). There are four peak structures present in the spectrum. The spin assignments of the three prominent peaks come from an analysis of the mirror, ${}^{12}$Be, presented in \cite{FortuneO12}, the predictions of \cite{Wang:2019,fortune2016O13removal, fortunesherr2011}, and previous work \cite{Jager:2012}. The low-energy shoulder on the  2$^{+}_1$ peak [Fig. \ref{fig:calibrationsandO12}(c)] is at the approximate location of the second 0$^{+}$, $E^{*}=1.62(13)$ state measured in \cite{Suzuki:2016}. It is also at the approximate energy for the 1$^-_1$ state predicted with GCC calculations ($E_T$=3.256~MeV) \cite{Wang:2019}.
 Alternatively, this structure may be a second 2$p$ decay branch of the 2$^+_2$ excited state  to the $E^{*}=3.353$-MeV, $J^{\pi}=2^{+}$, particle-bound first excited state as the shoulder is located $\sim$3.353~MeV lower in $E_{T}$ than the ground-state decay.

 The data were fitted with Monte-Carlo simulations using Breit-Wigner intrinsic lineshapes that incorporated the detector response. The numerics, extracted decay energies, widths, and cross sections of each state are given in Table \ref{tbl:O12}. The decay width of the shoulder state is subject to a large uncertainty, as demonstrated by the fit with different background parameterizations in Figs.~\ref{fig:calibrationsandO12}(c) and \ref{fig:calibrationsandO12}(d). The extracted centroid and width are consistent with the values of  $E^*$=1.62(13) and $\Gamma$=1.2~MeV  for the 0$^+_2$ state from \cite{Suzuki:2016}.
On the other hand,  if the shoulder peak is a decay branch of the second 2$^{+}$, we can fix its decay energy and width and refit the spectrum [Fig.~\ref{fig:2pB10}(a)], obtaining a branching ratio for this component of 0.27(1).

 This paper presents the first observation of a second 2$^{+}$ state, which was predicted to be produced in $^{13}$O knockout reactions in \cite{fortune2016O13removal}. Note that the analysis of \cite{fortune2016O13removal} predicts four 2$^{+}$ states, though only two are expected to have significant strength (the first and fourth) and the fourth (presumably our 2$^+_2$ state) should have four times the yield of the 2$^+_1$ state, which is not observed.

\begin{table*}
\caption{The spin assignments, decay energies, intrinsic widths, and measured cross sections  of the four observed states in ${}^{12}$O.}
\begin{ruledtabular}
\begin{tabular}{c c c c c }
$J_{\pi}$ & $E_{T}$ [MeV] & $E^{*}$ [MeV] & $\Gamma$ [MeV] & $\sigma$ [mb] \\
\hline 
$0^{+}_{1}$ & 1.718(15) & 0 & 0.051(19) & 1.2(2)  \\
$0^{+}_{2}$ \footnotemark[1] & 3.519(67) & 1.801(67) & 0.980(182) & 3.3(10) \\
$2^{+}_{1}$ & 3.817(18) & 2.099(18) & 0.155(15) & 2.1(5)  \\
$2^{+}_{2}$ & 6.493(17) & 4.775(17) & 0.754(25) & 3.5(9)  \\
$2^{+}_{2}$ \footnotemark[2] & 6.493(17) & 4.775(17) & 0.754(25) & 4.8(12)  \\
 & 7.972(40)\footnotemark[3] & 9.984(40) & 0.81(11) & 0.93(11) \\
\end{tabular}
\footnotetext[1]{Assuming the shoulder peak is the second 0$^{+}$.}
\footnotetext[2]{Assuming the shoulder peak is a decay branch of the second 2$^{+}$.}
\footnotetext[3]{Through the $4p+2\alpha$ decay channel.}
\end{ruledtabular}
\label{tbl:O12}
\end{table*}

\subsubsection{4$p$+2$\alpha$ channel}
\begin{figure}
\includegraphics[scale=0.4]{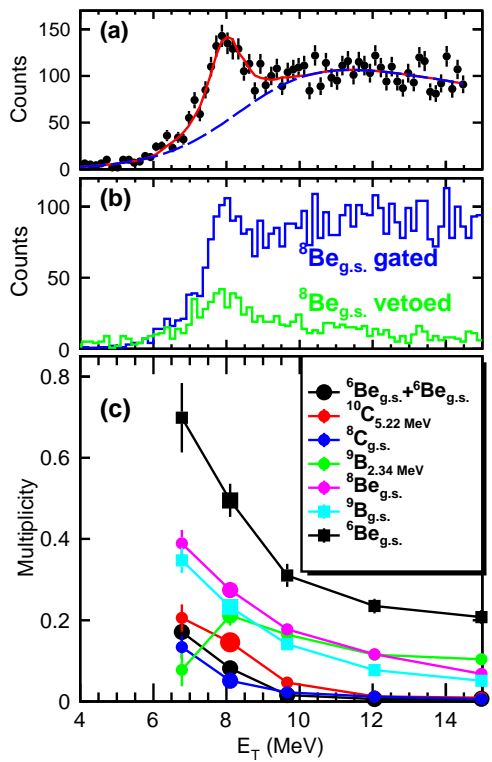}
\caption{(a) Total decay energy spectrum reconstructed from coincident 4\textit{p}+2$\alpha$ events. The red solid curve shows the result of a Monte-Carlo simulation with a Breit-Wigner intrinsic line shape. The blue dashed curve shows the fitted background. (b) Coincident 4\textit{p}+2$\alpha$ events with a gate on and a gate vetoing  $^{8}$Be$_{g.s.}$ intermediate states, and (c) the multiplicity of observed intermediate resonances assessed at five different points along the spectrum. Also included is the probability of observing a $^6$Be$_{g.s.}$+$^6$Be$_{g.s.}$ fission event. No $\cos\theta_{C}$ gate was employed.}
\label{fig:4p2a}
\end{figure}

Figure \ref{fig:4p2a}(a) displays the reconstructed total-decay-energy spectrum of detected 4$p$+2$\alpha$ events which shows a  peak sitting atop a large background. Assuming a single level with a Breit-Wigner line shape, the fitted spectrum is shown as the solid curve with fitted parameters: $E_T$=7.972(40)~MeV, $\Gamma$=0.81(11)~MeV. 

 With the large multiplicity of exit-channel fragments, it is not easy to disentangle how this state decays, especially as it sits on a large background. Among the possible decay scenarios include sequential decay pathways which are initiated by an $\alpha$+$^{8}$C/$^{6}$Be+$^{6}$Be split or 2$p$ decay to higher-lying excited states of $^{10}$C.  In all these scenarios, sequential decays of the unstable decay products  lead to the observed exit channel.
 
Of the possible  intermediate states, $^8$Be$_{g.s.}$ is cleanest to isolate as there is only one possible $\alpha$-$\alpha$ subevent in each 4$p$+2$\alpha$ event and the $\alpha$-$\alpha$ invariant-mass peak has no background under it. The invariant-mass spectrum gated and vetoed by the presence of a $^8$Be intermediate state is shown in Fig.~\ref{fig:4p2a}(b). Peaks are present in both spectra, indicating there are at least two decay pathways, but, as we will see, there are many more.

\begin{figure}
\includegraphics[scale=0.5]{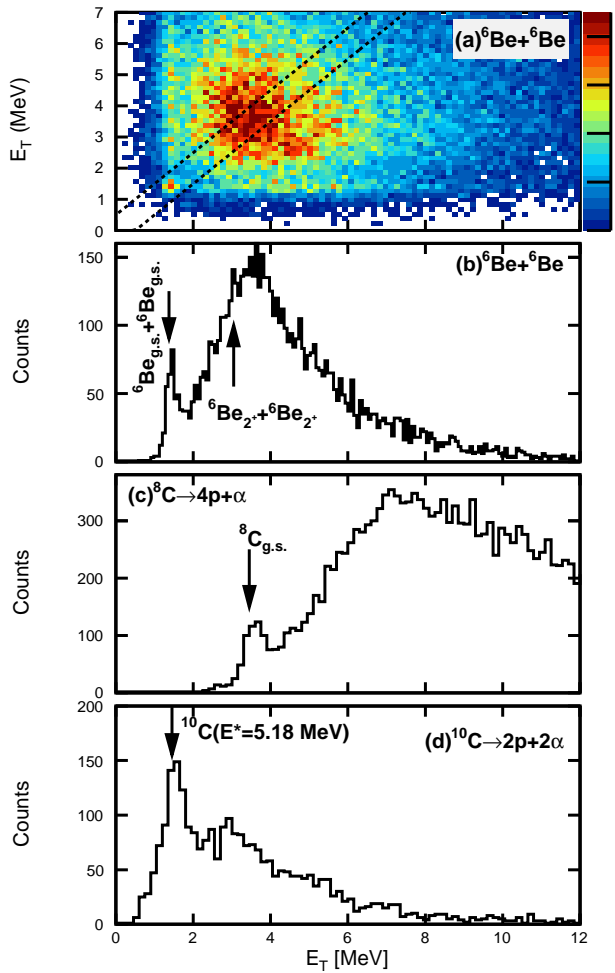}
\caption{Invariant-mass spectra showing intermediate states present in the 4$p$+2$\alpha$ channel. (a) shows the correlation of the $E_T$ values of  two $^6$Be's that can be reconstructed from the six detected particles. If the diagonal gate indicated by the dotted lines is applied, the projection on either one of the axes is shown in (b).  (c) and (d) show invariant-mass spectra for 4$p$+$\alpha$ and 2$p$+2$\alpha$ subevents. The arrows label the peaks associated with known levels.  No $\cos\theta_{C}$ gate was employed.}
\label{fig:4p2achannels}
\end{figure}

Other possible intermediate states can be identified using various combinations of the decay fragments. For instance, the invariant-mass spectra of 4$p$+$\alpha$ subevents is shown in Fig.~\ref{fig:4p2achannels}(c). Note there are two ways to make  4$p$+$\alpha$ subevents for each 4$p$+2$\alpha$ event. This spectrum shows the presence of a peak from the decay of  $^8$C$_{g.s.}$ at $E_T$=3.449~MeV \cite{Charity:2011}. On the other hand, for $^{10}$C intermediate states, the combinatorics give us 6 2$p$+2$\alpha$ subevents for each event. Rather than dealing with all these subevents, the invariant-mass spectrum  in Fig.~\ref{fig:4p2achannels}(d) used only 2$p$+2$\alpha$ subevents where one of the $p$+2$\alpha$ subsubevents has an invariant mass corresponding to $^9$B$_{g.s.}$ decay. This spectrum shows the presence of a prominent peak at $E_T$=1.45~MeV corresponding to the 5.18-MeV state of $^{10}$C which is known to sequentially proton decay through $^9$B$_{g.s.}$ to $^8$Be$_{g.s.}$ \cite{Charity:2007}. Finally there are also 6 ways to separate a detected event into two subgroups of 2$p$+$\alpha$ events from which one can search for the fission of $^{12}$O into two $^6$Be fragments. Plotting the decay energies of the  prospective $^6$Be fragments against each other produces the 2D plot shown in Fig.~\ref{fig:4p2achannels}(a). A sharp peak for which the two $^6$Be fragments are in their ground states can be discerned ($E_T$=1.37~MeV). To make this more obvious, we have employed the diagonal gate shown by the dotted lines and projected onto either one of the axes to produce the spectrum shown in Fig.~\ref{fig:4p2achannels}(b). Apart from g.s.+g.s. peak there is a broad peak at $\sim$3.5~MeV somewhat above the  expected energy if both $^6$Be fragments are in the $E^*$=1.16~MeV, $J^\pi$=2$^+$ state. Its origin is not clear. 

By subtracting a smooth background under all the peaks in Fig.~\ref{fig:4p2achannels}, a multiplicity for each observed intermediate resonance and a probability for the $^6$Be$_{g.s.}$+$^6$Be$_{g.s.}$ channel can be determined. These various multiplicities are plotted against $E_T$ in Fig.~\ref{fig:4p2a}(c). In addition to the resonances discussed, multiplicities were also determined for the occurrence of a  single $^6$Be$_{g.s.}$ and for the  $J^\pi$=5/2$^-$, $E^*$=2.345-MeV level of $^9$B. These multiplicities are inclusive and so for example the $^8$Be$_{g.s.}$ yield from the decay of $^9$B$_{g.s.}$ is included in the $^8$Be$_{g.s.}$ multiplicity. The $E_T$ bin centered around the observed peak is marked by the fatter data points near 8 MeV. If these intermediate resonances are associated mainly with the $^{12}$O peak and not with the large background under it, then the multiplicities should have a maximum at the $E_T$ value for the peak ($E_T^{peak}$). This is clearly not the case. On the other hand, if the intermediate resonances come purely from the background, then the multiplicities should have a minimum at $E_T^{peak}$. According to our fit in Fig.~\ref{fig:4p2a}(a), the peak to background yield for the $E_T$ bin centered on the peak is $\sim$1:1. Thus we would expect a suppression of 50\% at the peak $E_T$ relative to the trend from the other $E_T$ bins in this scenario. Again this is also not the case. Thus we conclude that all these intermediate resonances are found in both the peak and the background beneath it with similar magnitudes. Making a more quantitative statement is impossible as the overall trend is non-linear and thus difficult to determine if the peak multiplicities are slightly larger or smaller than the overall trend. 

We conclude that about 50\% of the events in the observed $^{12}$O peak decay by producing at least one $^6$Be$_{g.s.}$ and another $\sim$20\% have a $^9$B$_{g.s}$ fragment. The latter is responsible for most of the observed $^8$Be$_{g.s.}$ yield.  The fission-like channels $^6$Be$_{g.s.}$+$^6$Be$_{g.s.}$ and $\alpha$+$^8$C$_{g.s.}$ are responsible for $\sim$9\% and $\sim$6\% of the yield, respectively.  The above analysis may miss wider intermediate states which are more difficult to discern above the backgrounds.      

\subsection{$^{12}$N structure}
\subsubsection{2$p$+$^{10}$B channel}
\begin{figure}
\includegraphics[scale=0.4]{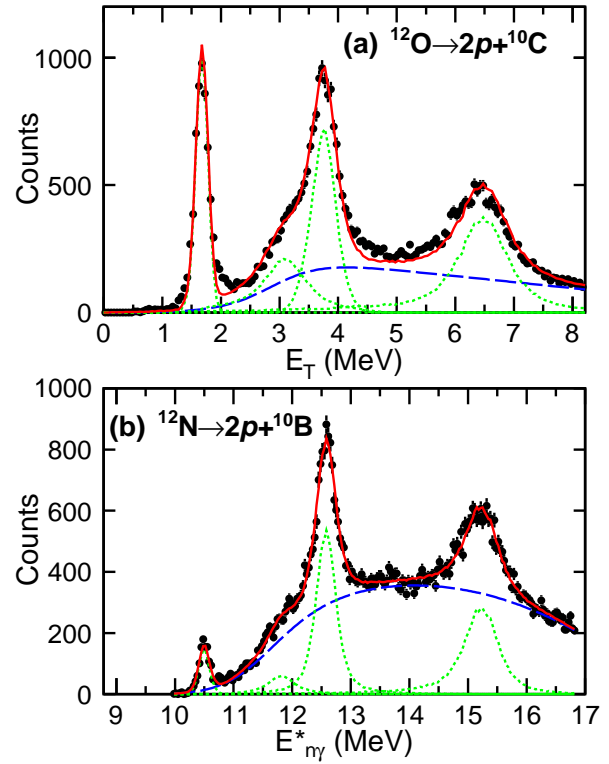}
\caption{Comparison of invariant-mass spectra for the (a) 2$p$+$^{10}$C and (b) 2$p$+$^{10}$B exit channels. Fits using Breit-Wigner intrinsic line shapes are shown by the curves in both panels. A $\left|\cos\theta_{C}\right|<$0.2 gate was applied for both spectra.}
\label{fig:2pB10}
\end{figure}

Known low-lying $^{12}$N states were observed in the $p$+$^{11}$C spectrum in Fig.~\ref{fig:calibrationsandO12}(b). Higher-lying excited states can be found in the 2$p$+$^{10}$B spectrum in Fig.~\ref{fig:2pB10}(b). This spectrum bears a strong resemblance to the 2$p$+$^{10}$C spectrum which is also reproduced in panel (a) for convenience.   Both spectra contain three prominent peaks with the same energy spacing between them. In addition, the second of these peaks in both cases has a low-energy shoulder. This surprising result suggests that all the peaks in the 2$p$+$^{10}$B spectrum  are analogs of the observed $^{12}$O levels. Indeed, the lowest-energy peak was also  observed in the invariant-mass study of Jager \textit{et al.} \cite{Jager:2012} and it was argued that it was produced by the prompt 2$p$ decay of the Isobaric Analog State (IAS) of $^{12}$O in $^{12}$N  to the IAS in $^{10}$B. The latter subsequently $\gamma$ decays. This represented a second example of prompt 2$p$ decay from one IAS to another \cite{Charity:2011}. The other example is the IAS of $^{8}$C in $^8$B which 2$p$ decays to the IAS state in $^6$Li where the $\gamma$ ray from the decay of $^6$Li$_{IAS}$ was observed \cite{Brown:2014}.  In both cases, prompt 2$p$ decay was the only particle-decay mode which conserved both energy and isospin. 

\begin{table*}
\caption{Fitted parameters for $^{12}$N states observed in the 2$p$+$^{10}$B invariant-mass spectrum of Fig.~\ref{fig:2pB10}(b), the 2$p$+$\alpha$+$^6$Li spectra in Figs.~\ref{fig:2paLi6}(c) and \ref{fig:2paLi6}(d), and the 3$p$+$^9$Be spectrum in Fig.~\ref{fig:3pBe9}. } 
\label{tbl:N12}
\begin{ruledtabular}
\begin{tabular}{c c c c c c c}
Observed & Intermediate & $J^{\pi}$ & $E^{*}_{n\gamma} [MeV]$ & $E^*$ [MeV] &  $\Gamma$ [keV] & $\sigma$ (mb)\\
Channel  &  Channel     &           &       &  &   & \\
\hline 
2$p$+$^{10}$B  & & 0$^+$ & 10.502(4) & 12.242 (4)  & $<$100 & 0.19(6)\\
2$p$+$^{10}$B  & & 2$^+$ & 12.574(4) & 14.314 (4)  & 193(26) & 1.2(3)\\
2$p$+$^{10}$B & & 2$^+$ & 15.252(11) & 16.992(11) & 420(120) & 2.18(5)\footnotemark[1]\\ 
2$p$+$\alpha$+$^6$Li  & $\alpha$+$^{8}$B$_{IAS}$ &  & 18.408(66) &21.970(66) & 649(49) & 0.091(2) \\                    
2$p$+$\alpha$+$^6$Li  & $^{6}$Be$_{g.s.}$+$^{6}$Li$_{IAS}$ &  & 18.569(86) & 22.131(86)&  702(170)   & 0.15(4)      \\
3$p$+$^9$Be &      &  &  22.158(23) & 22.158(23) & 138(79) & 2.5(6)                                                    \\
\end{tabular}
\end{ruledtabular}
\footnotetext[1]{Includes contribution from the shoulder state.}
\end{table*}

If these other states are also analog states ($T$=2), then the two higher-energy states are both $J^\pi$=2$^+$ and  also involve 2$p$ decay to the IAS of $^{10}$B.  The low-energy shoulder on the lowest 2$^+$ peak can again have contributions from three possible sources, i.e. from $T$=2, $J^\pi$=0$^{+}$ or 1$^{-}$ levels or from a decay branch of a higher-energy 2$^+$ state which decays to the first $T$=1, $J^\pi$=2$^+$ state in $^{10}$B (Fig.~\ref{fig:level_2pB10}). The red solid curve in Fig.~\ref{fig:2pB10}(b) shows a fit obtained with Breit-Wigner intrinsic line shapes where the third origin of the low-energy shoulder is assumed. Figure \ref{fig:2pB10}(a) shows a similar fit to the 2$p$+$^{10}$C decay channel of $^{12}$O, assuming the shoulder state decays to the excited state of $^{10}$C. For the decay of $^{12}$N, the known energy of the first  $T$=1, $J^\pi$=2$^+$ $^{10}$B state is used to constrain its $E^*_{n\gamma}$ centroid relative to the fitted value for the second 2$^+$ state. The fitted branching ratio is 0.087(12) which is smaller than the corresponding value obtained for the $^{12}$O$\rightarrow$2$p$+$^{10}$C shoulder state (see Sec.~\ref{sec:O12}).

Although the 2$p$+$^{10}$B and 2$p$+$^{10}$C spectra in Fig.~\ref{fig:2pB10} have similar structure, the obvious difference is the much smaller relative yield for the $^{12}$N$_{IAS}$ peak ($E^*_{n\gamma}$=10.502~MeV) compared to that for the other peaks.  The cross sections for the 2$^+_1$ and 2$^+_2$ levels of $^{12}$N are about $\sim$50\% of the corresponding $^{12}$O levels. Based on this scaling, we expect the yield of  0$^+$ $^{12}$N IAS to be a factor of $\sim$3 larger than observed.
  The simplest explanation would be that  $^{12}$N$_{IAS}$ has one or more other decay branches which account for most of the decay leaving the 2$p$ branch with only $\sim$30\% of the yield. If these other branches involve particle emission, then they must violate isospin symmetry, but no candidate peaks were observed in other $^{11}$N exit-channels. We cannot rule-out other particle-decay branches as some of these will have very low detection efficiency, e.g., single-proton decay to the $^{11}$C$_{g.s.}$ where the emitted proton will mostly travel to angles outside the detectors angular acceptance.  Another possibility is that the $^{12}$N$_{IAS}$ has a significant M1 $\gamma$ decay strength to  the ground state which would satisfy the spin and isospin selection rules.

\subsubsection{2$p$+$\alpha$+$^{6}$Li channel}

\begin{figure}
\includegraphics[scale=0.4]{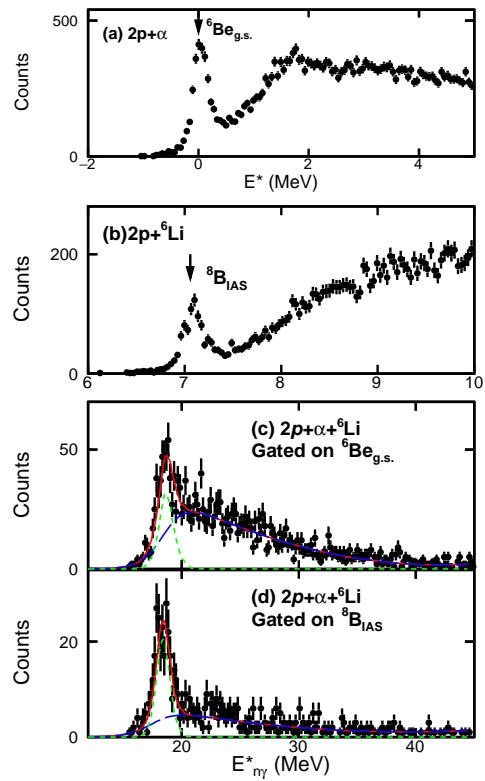}
\caption{Data obtained for 2$p$+$\alpha$+$^6$Li coincidences. 
The upper panels show  (a) 2$p$+$\alpha$ and  (b) 2$p$+$^6$Li invariant-mass spectra for these subevents. No $\cos\theta_C$ gate was applied. (c,d) show invariant-mass spectra obtained from the 2$p$+$\alpha$+$^6$Li events gated on the $^6$Be$_{g.s.}$ and $^8$B$_{IAS}$ invariant-mass peaks in (a) and (b), respectively. The curves in (c,d) are fits assuming a single Breit-Wigner lineshape.
}   
\label{fig:2paLi6}
\end{figure}

While the analogs of the low-lying $^{12}$O levels have been found, we have also searched for the analog of the higher-lying $^{12}$O level observed in the 4$p$+2$\alpha$ channel. Given the large number of decay paths inferred for this state, it is reasonable to assume multiple exit-channels are associated with the analog state, some of which would contain neutrons and not accessible with the experimental apparatus. However, the analog of the fission-like $^6$Be$_{g.s.}$+$^6$Be$_{g.s}$ and $\alpha$+$^8$C$_{g.s.}$ channels in $^{12}$O would be $^6$Be$_{g.s.}$+$^{6}$Li$_{IAS}$ and $\alpha$+$^8$B$_{IAS}$  which would populate the 2$p$+$\alpha$+$^6$Li+$\gamma$ exit channel. No peaks are observed in the raw 2$p$+$\alpha$+$^6$Li invariant-mass spectrum, but with suitable gating the analogs can be found. The invariant-mass spectra from  the 2$p$+$\alpha$ and  2$p$+$^6$Li subevents are shown in Fig.~\ref{fig:2paLi6}(a) and \ref{fig:2paLi6}(b), respectively. 

The peak observed in the 2$p$+$^6$Li exit channel [Fig.~\ref{fig:2paLi6}(b)] is consistent in energy and width to a peak observed following proton knockout from a $^9$C beam \cite{Brown:2014}. It was identified as a  $^8$B$_{IAS}\rightarrow$2$p$+$^6$Li$_{IAS}$ decay where the $\gamma$ ray from the decay of $^6$Li$_{IAS}$ was observed in a subsequent experiment. Gating on the $^{8}$B$_{IAS}$ peak produces the 2$p$+$\alpha$+$^6$Li invariant-mass spectrum in Fig.~\ref{fig:2paLi6}(d). Taking into account the energy of the unobserved gamma ray (3.563~MeV), the excitation energy of the $^{12}$N$\rightarrow\alpha$+$^8$B$_{IAS}$ state is at 21.970(66)~MeV. 

A $^6$Be ground-state peak is observed in the invariant-mass spectrum for the 2$p$+$\alpha$ subevents [Fig.~\ref{fig:2paLi6}(a)]. Gating on this peak produces the invariant-mass spectrum in Fig.~\ref{fig:2paLi6}(c). In principle this peak could correspond to either a $^6$Be$_{g.s.}$+$^6$Li$_{g.s.}$ or $^6$Be$_{g.s.}$+$^6$Li$_{IAS}$ decay. However, the fitted $E^*_{n\gamma}$ values and widths of the peaks in Figs.~\ref{fig:2paLi6}(c) and \ref{fig:2paLi6}(d) (given in Table~\ref{tbl:N12}) are consistent, suggesting that they are decay branches of the same state. Thus the peak in Fig.~\ref{fig:2paLi6}(c) corresponds to the fission-like $^6$Be$_{g.s.}$+$^6$Li$_{IAS}$ decay.

\subsubsection{3$p$+$^9$Be channel}

\begin{figure}
\includegraphics[scale=0.4]{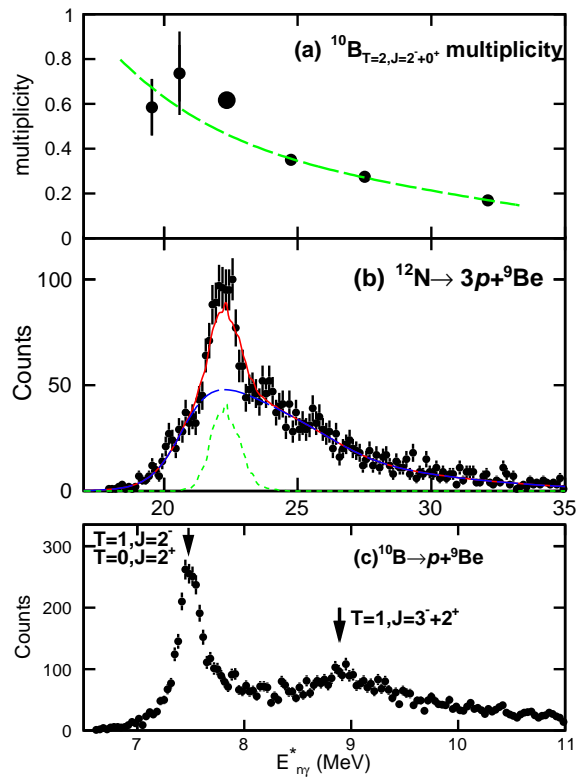}
\caption{(b) Experimental  3$p$+$^9$Be invariant-mass spectrum with a fit assuming a single  Breit-Wigner level and a smooth background (blue dashed curve).
(c) Invariant-mass spectrum obtained from the $p$+$^9$Be subevents of these events. The energies of known $^{10}$B levels are indicated. (a) The multiplicity associated with the low-energy peak in (c) as a function of the $^{12}$N excitation energy. The dashed blue curve is a smooth fit to these multiplicities excluding the fatter data point  at the $^{12}$N peak energy.  A  $\left|\cos\theta_{C}\right|<$0.5 gate was applied to both spectra. }
\label{fig:3pBe9}
\end{figure}

As the $^{12}$O$\rightarrow$4$p$+2$\alpha$ peak appears to have a significant probability of decaying to $^9$B$_{g.s.}$ intermediate states, its analog in $^{12}$N could decay to the 3$p$+$^{9}$Be exit channel. The invariant-mass spectrum for these events is shown in Fig.~\ref{fig:3pBe9}(b) and displays a well resolved peak at an excitation energy consistent with that inferred for the $^{12}$N$\rightarrow$2$p$+$\alpha$+$^6$Li peaks. See Table~\ref{tbl:N12} for comparison of peak parameters from Breit-Wigner fits. However the fitted intrinsic width of this state is considerably smaller than the ($T$=2)  2$p$+$\alpha$+$^6$Li peaks, indicating that it must be associated with a different $^{12}$N level. It may even be a $T$=1 state. However we suspect it is a $T$=2 state, and if this is true, it implies that the observed peak in the analog spectrum, $^{12}$O$\rightarrow$4$p$+2$\alpha$, is not a singlet. 

Further information on the decay of this state can be gleaned from the invariant-mass spectrum of the $p$+$^{9}$Be subevents in Fig.~\ref{fig:3pBe9}(c). A prominent peak at $E^*\sim$7.47~MeV is observed which could be one or both of the $E^*$=7.470-MeV ($J^\pi$=2$^+$, $T$=0) and $E^*$=7.479-MeV ($J^\pi$=2$^-$, $T$=1) excited states of $^{10}$B.  The multiplicity of this peak is plotted as a function of $^{12}$N excitation energy in Fig.~\ref{fig:3pBe9}(a). The 
smooth green dashed curve was fitted to excitation-energy bins beyond the peak region and suggests that there might be a small enhancement of this intermediate state at the peak energy. However, this curve is not well constrained for energies below the peak value. Nevertheless, it is clear that this intermediate-state peak is associated with a significant fraction of the $^{12}$N$\rightarrow$3$p$+$^9$Be peak yield. If the $^{12}$N state is $T$=2 as suspected, then the intermediate state must be the $T$=1, $J^\pi$=2$^-$ state. We also see a suggestion of some smaller yield for the $T$=1, $J^\pi$=3$^-$($E^*$=8.887~MeV) and 2$^+$ ($E^*$=8.895~MeV) $^{10}$B states in the $p$+$^{9}$Be spectrum in Fig.~\ref{fig:3pBe9}(c), but it is difficult to extract multiplicities as the  relative magnitude of the background is much larger.

\section{Discussion}
\subsection{Analog States}

\begin{figure}
\includegraphics[scale=0.4]{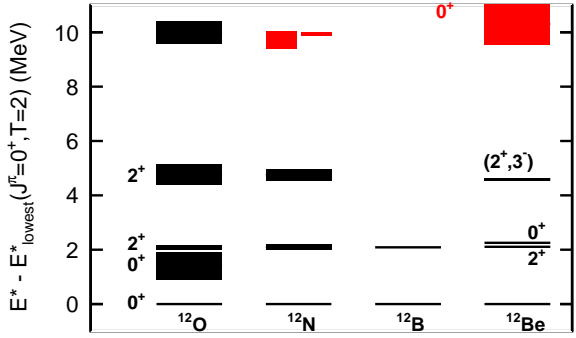}
\caption{Level schemes for $A=12$ and $T$=2 states with excitation energy plotted relative to the first $J^\pi$=0$^+$ state.
The results for $^{12}$O and $^{12}$N are from this work, the $^{12}$B levels are from \cite{Charity:2008}, and the $^{12}$Be levels from \cite{ENSDF, Yang:2015}.}
\label{fig:analog}
\end{figure}

To further explore the analog states in $^{12}$O and $^{12}$N, we have looked at their consistency with the other $A$=12 quintet states. Figure~\ref{fig:analog} plots the level scheme of the newly found levels relative to  the energy of the lowest $T$=2, $J^\pi$=0$^+$ state. In addition, $T$=2 levels for $^{12}$B \cite{Charity:2008} and $^{12}$Be \cite{ENSDF} are included.   There is excellent consistency for the energy of the first $T$=2, $J^\pi$=2$^+$ state. The second such 2$^+$ state in $^{12}$Be is not known, but a state at $E^*$=4.580 MeV has been assigned as either 2$^+$ or 3$^-$ \cite{ENSDF}. This state is at the right energy to be the mirror of the 2$^+_2$ state in $^{12}$O and thus leads us to strongly favor the 2$^+$ assignment.

\begin{figure}
\includegraphics[scale=0.4]{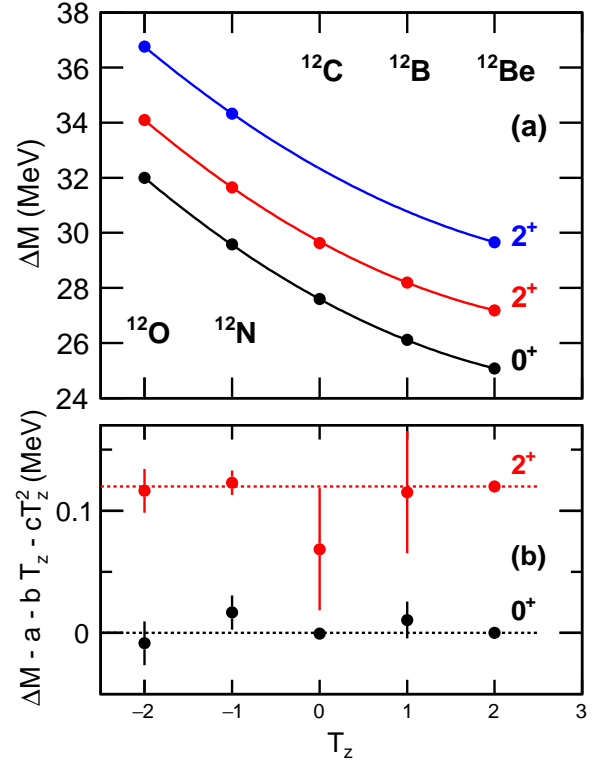}
\caption{(a) Mass excesses of three lowest $A$=12 quintets as a function of isospin projection $T_{Z}$. The curve through the data points are fits with the quadratic IMME [Eq.~(\ref{eq:IMME})]. (b) Deviations from the IMME fit for the lowest quintets. The results for the 2$^+$ states have been shifted up along the $y$-axis for clarity. The statistical and systematic errors for the $^{12}$O and $^{12}$N data points from this work are of similar magnitude. We have combined them in quadrature for these fits.}
\label{fig:IMME}
\end{figure}

The masses in an isospin multiplet are expected to follow the isospin multiplet mass equation (IMME),
\begin{equation}
\Delta M = a + b T_{z} + c T_{Z}^2,
\label{eq:IMME}
\end{equation} 
if isospin is a good quantum number \cite{Benenson:1979}. The largest deviations from this behavior have been found for $A$=8 and 9 \cite{Benenson:1979,Lam:2013,Charity:2011}. Deviations from the  IMME for $A$=12 quintet were previously examined in \cite{Jager:2012}, but with our newer, more accurate measurement for $^{12}$O and $^{12}$N it is worth examining again. The mass excess for the quintets associated with the 0$^+$ and first and second 2$^+$ $T$=2 states are plotted in Fig.~\ref{fig:IMME}(a). Results for $^{12}$C were taken from \cite{ENSDF} and  for the second 2$^+$ state of $^{12}$Be, we have included the candidate state previously mentioned. The curves show fits with the quadratic IMME which can reproduce the data validating our assumption that the $^{12}$N states are $T$=2.

Figure~\ref{fig:IMME}(b) shows deviations from the quadratic form for the two lower-energy quintets. The error bars on some of the points for the 2$^+$ states are too large to look for meaningful deviations from the IMME. In the case of the 0$^+$ states, if we include a cubic term ($d T_{z}^3$) to Eq.~(\ref{eq:IMME}), then we find $d$ = 1.6(37)~keV consistent with zero and small compared to values of $d$=8.36(21)~keV for the $A$=8 quintet and $d$=6.33(16)~keV for the $A=9$ quartet \cite{Britz:1998}. There is thus no evidence for deviations from the quadratic IMME for the $A$=12 quintet. 

The fission-like channels $^6$Be$_{g.s.}$+$^6$Be$_{g.s.}$, $^6$Be$_{g.s.}$+$^{6}$Li$_{IAS}$, $\alpha$+$^8$C$_{g.s.}$, and $\alpha$+$^8$B$_{IAS}$  remind us of the $^6$He+$^6$He and $\alpha$+$^8$He decay channels for levels in $^{12}$Be observed in \cite{Freer:1999, Charity:2007a} and which were associated with a molecular $\alpha$:2$n$:$\alpha$ band structure \cite{Freer:1999}. The lowest member of this band was subsequently found at $E^*$=10.3~MeV \cite{Yang:2014,Yang:2015} and shown to be $J^\pi$=0$^+$. The location of this level is plotted in Fig.~\ref{fig:analog} for comparison. While its excitation energy would be appropriate for an analog to the fission-like states we observed in $^{12}$O and $^{12}$N, its quoted width of $\Gamma$ = 1.5(2)~MeV is too large to be consistent as more neutron-rich analogs are closer to stability and thus should have smaller widths. In any case these fission-like decay modes suggest these states have strong cluster structure.

\subsection{Momentum Correlations}
\label{sec:corr}
The momentum correlations in a 2$p$ decay can be presented as a 2D distribution \cite{Pfutzner:2012} and are typically given as an energy ($E_x/E_T$) versus angle 
[$\cos\theta_k= \mathbf{p}_x\cdot\mathbf{p}_y/(p_x p_y)$] plot  where 
\begin{gather}
 E_{T} = E_x + E_y = \frac{(A_1+A_2)p_x^2}{2M A_1 A_2}  + \frac{(A_1+A_2+A_3)p_y^2}{2 M (A_1+A_2)A_3}, \\
\mathbf{p}_x = \frac{A_2 \mathbf{p}_{1} - A_1\mathbf{p}_{2}}{A_1 + A_2}, \\
\mathbf{p}_y = \frac{A_3(\mathbf{p}_1+\mathbf{p}_2) - (A_1+A_2)\mathbf{p}_3}{A_1+A_2+A_3},
\end{gather}
and the mass of the parent is $M=(A_1+A_2+A_3)m_0$, where $m_0$ in the nucleon mass  and $\mathbf{p}_1$, $\mathbf{p}_2$, and $\mathbf{p}_3$ are the momenta of the decay products. 

\begin{figure*}
\includegraphics[scale=0.7]{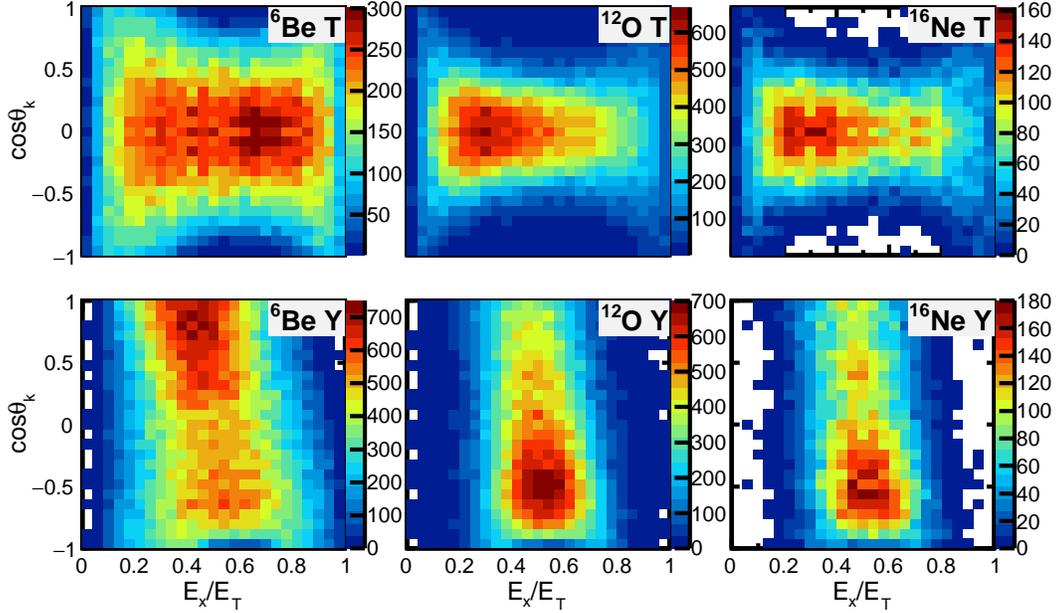}
\caption{Experimental momentum correlation plots for the ground-state 2$p$ decays of $^6$Be, $^{12}$O, and $^{16}$Ne. Results are shown in both the Jacobi T (upper panels) and Jacobi Y (lower panels) representations. The $^{12}$O data is from this work while the $^6$Be and $^{16}$Ne data are from Refs.~\cite{Egorova:2012,Brown:2015}.}
\label{fig:Jacobi2d}
\end{figure*}

If $\mathbf{p}_3$ is the core momentum, then the $x$ and $y$ quantities are associated with the Jacobi T representation, otherwise  we get the Jacobi Y representation.  The quantity $E_x$ in the Jacobi T representation is the relative energy between the protons ($E_x=E_{pp}$), while in the Jacobi Y representation it is the relative energy between the core and one of the protons ($E_{x}=E_{p-core}$). In the Jacobi-Y case, $\cos\theta_k\sim$-1 corresponds to a small relative angle between the momenta of the two protons. The Jacobi 2D plots for $^{12}$O$_{g.s.}$ are compared to those obtained for ground-state emission of $^6$Be \cite{Egorova:2012} and $^{16}$Ne \cite{Brown:2015} in Fig.~\ref{fig:Jacobi2d}.

 \begin{figure}
\includegraphics[scale=0.4]{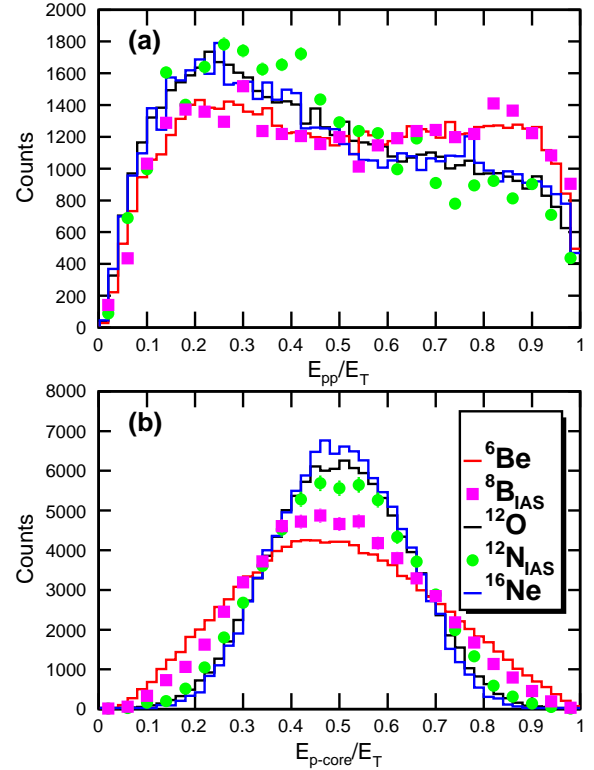}
\caption{Projections of the (a) Jacobi T and (b) Jacobi Y 2D distributions on the energy axis. Ground-state 2$p$ emitters are plotted as histograms while 2$p$ emission from isobaric analogs is plotted as the data points.}
\label{fig:Jacobi1d}
\end{figure}

All these data were obtained with the same detector setup, and approximately the same beam $E/A$ value, so the effect of the detector resolution and  efficiency will be similar. All channels display features that are expected for 2$p$ decay, including suppressions due to the Coulomb interaction when one of the protons and the core have the same velocity vectors (e.g. $E_x/E_T\sim$0.56, $\cos\theta_k=\pm$1 in the Jacobi T plots) and  when the two protons have the same velocity vectors (e.q. $E_x/E_T\sim$0.5, $\cos\theta_k$=-1 in the Jacobi Y plots). It is also clear that the two protons are emitted with approximately the same kinetic energy, as the ridges in the Jacobi Y distribution are centered near $E_x/E_T\sim$0.5. 

 The $^{12}$O results are remarkably similar to those for $^{16}$Ne$_{g.s.}$, and both differ from the $^{6}$Be results in  having relatively more yield for diproton-like decays, i.e., small $E_x/E_T$ in the Jacobi T plots or negative $\cos\theta_k$ values in the Jacobi Y plots. Further comparison can be made with the projections of the distributions on their energy axis  which are  shown in Fig.~\ref{fig:Jacobi1d} as the histograms. The similarity of the $^{12}$O$_{g.s.}$ and $^{16}$Ne$_{g.s.}$ data are confirmed in these projections. This similarity was predicted by Grigorenko \textit{et al.} in their 3-body calculations \cite{Grigorenko:2002}.

In addition to the ground-state emitters, we have also included projections for the 2$p$ decays of $^{12}$N$_{IAS}$ and $^{8}$B$_{IAS}$ \cite{Brown:2014} as the data points.
We note all the correlation data can be subdivided into two groups with similar projections: the first group - lighter $p$-shell
systems $^6$Be$_{g.s.}$ and $^8$B$_{IAS}$ - and the second group - the heavier $^{12}$O$_{g.s.}$, $^{12}$N$_{IAS}$, and $^{16}$Ne$_{g.s.}$ systems. The $^8$B$_{IAS}$ 2$p$ decay may be considered a surrogate for the first step of the decay of its analog, i.e., $^8$C$_{g.s.}\rightarrow$2$p$+$^6$Be$_{g.s.}\rightarrow$4$p$+$\alpha$ \cite{Charity:2011}. One should also note the similarity of the correlations for $^{12}$N$_{IAS}$ to its analog $^{12}$O$_{g.s.}$ reinforcing this notion.  The degree of similarity is probably helped by the fact that $E_T$ is very similar for all systems (Table~\ref{tbl:2pemitters}), as the three body-model of Grigorenko \textit{et al.} indicates that there is a $E_T$ dependence of these projections \cite{Grigorenko:2003,Grigorenko:2009}.

The similarity for all the heavier systems is expected as they all have significant $[s_{1/2}^2]_0$ components in their wavefunction and this component dominates the decay. On the other hand, $^6$Be$_{g.s.}$ is dominated by the $[p_{3/2}^2]_0$ component, but there is some mixing with  $[s_{1/2}^2]_0$ under the barrier \cite{Grigorenko:2003,Grigorenko:2009} and the final distribution is a mixture of $[p_{3/2}^2]_0$ and $[s_{1/2}^2]_0$ components. The similarity between the projections for $^6$Be$_{g.s.}$ and $^8$B$_{IAS}$ ($^8$C$_{g.s.}$) is quite interesting.  The interior $^8$C wavefunction of the valence protons can contain $p_{3/2}^4$ and $p_{3/2}^2 p_{1/2}^2$ components. The emission of a $[p_{1/2}^2]_0$ pair of protons would be expected to yield correlations more like the heavier systems as $[p_{1/2}^2]_0$ emission is expected to have a similar phase space as $[s_{1/2}^2]_0$ \cite{Sharov:2019}. It is possible that the $^8$B$_{IAS}$  result is telling us that the wavefunction is dominated by the $p_{3/2}^4$ component as predicted in the complex scaling model \cite{Myo:2012}. However, other models predict the addition of more  complex configurations in the mirror nucleus \cite{Papadimitriou:2011}. Calculations of the correlations for this system would help us better understand the structure of $^8$C, but the systematics of the presented correlations for light 2$p$ emission demonstrate that they contain structural information for the parent state.

\begin{table}
\caption{Decay energies for prompt 2$p$ emitters in Fig.~\ref{fig:Jacobi1d}.}
\label{tbl:2pemitters}
\begin{ruledtabular}
\begin{tabular}{ccc}
Nucleus & $E_{T}$ [MeV] & Ref. \\
\hline
$^6$Be & 1.372 & \cite{AME2016}\\
$^8$B$_{IAS}$ & 1.313 & \cite{AME2016}\\
$^{12}$O$_{g.s.}$ & 1.718(29) & this work \\
$^{12}$N$_{IAS}$ & 1.165(29) & this work \\
$^{16}$Ne$_{g.s.}$ & 1.466(20) & \cite{Brown:2015}\\
\end{tabular}
\end{ruledtabular}
\end{table}

\begin{figure}

\includegraphics[scale=0.4]{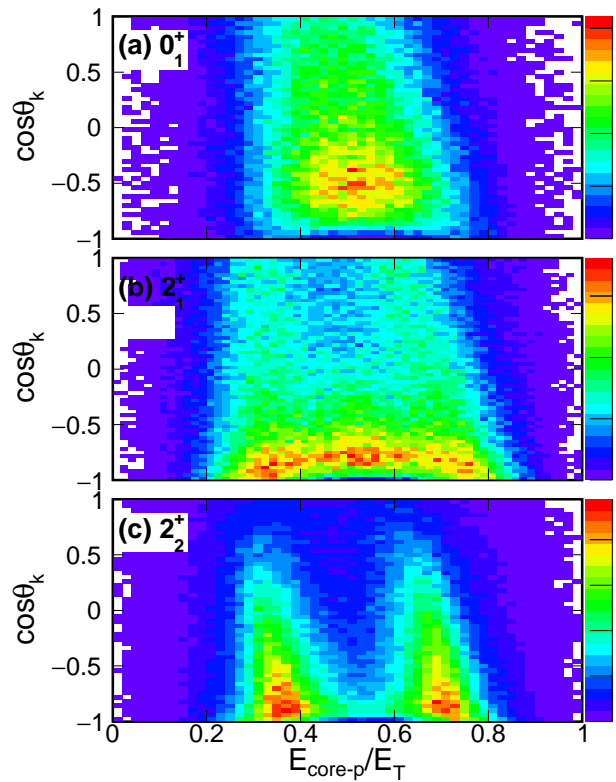}
\caption{Jacobi Y momentum correlations for $E_T$ bins centered on (a) 0$^+_1$, (b) 2$^+_1$, and (c) 2$^+_2$ states. The 2$^+_1$ results in (b) contained contributions from the shoulder state. The energy gates for the spectra are $1.43<E_{T}<2.02$ MeV for (a), $3.45<E_{T}<4.04$ MeV for (a), and $5.85<E_{T}<6.06$ MeV for (c).}
\label{fig:O12jacobis}
\end{figure}

Figure \ref{fig:O12jacobis} shows the Jacobi Y momentum correlations for $E_T$ bins centered on the three prominent peaks (0$^+_1$, 2$^+_1$, and 2$^+_2$) in the 2$p$+$^{10}$C invariant-mass spectra in Fig.~\ref{fig:calibrationsandO12}(c).  The 2$^+_1$ bin contains contributions from the shoulder state and the 2$^+_2$ bin contains a background contribution. Examination of the neighboring regions to these peaks suggest these contributions have similar correlations.  The correlations evolve from the prompt 2$p$ decay for the ground state [Fig.~\ref{fig:O12jacobis}(a)] to a sequential-like decay for the 2$^+_2$ state which displays  two vertical ridges [Fig.~\ref{fig:O12jacobis}(c)].
The latter decay is more complicated than the classical notion of sequential as we will see, but is it useful to determine which $^{11}$N intermediate state these ridges are consistent with. 

\begin{figure}
\includegraphics[scale=0.4]{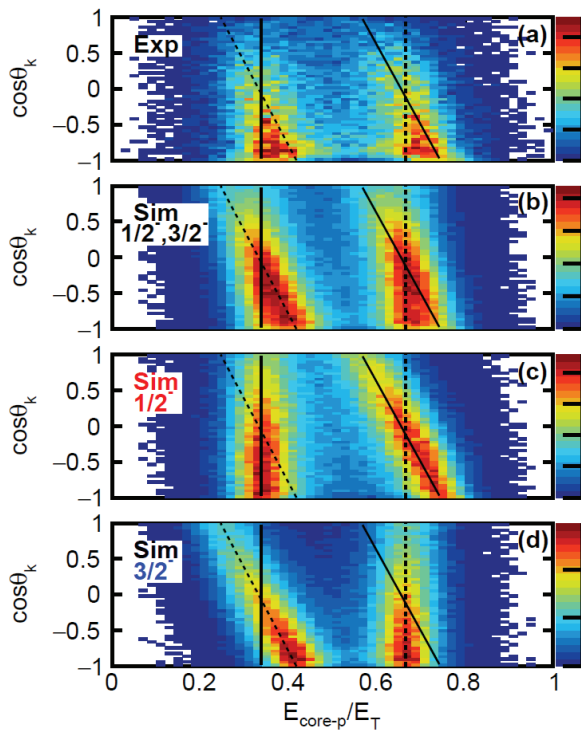}
\caption{(a) Experimental 2D momentum correlations of the 2$^{+}_{2}$ state of $^{12}$O, and simulated momentum correlations of sequential decay through (b) a combination of the $\frac{1}{2}^{-}$ and $\frac{3}{2}^{-}$ second and fifth excited states of $^{11}$N, (c) the $\frac{1}{2}^{-}$ states alone, and (d) the $\frac{3}{2}^{-}$ state alone. The decay scheme illustrating these states is shown in Fig. \ref{fig:level_2pB10}. The dashed lines show the expected ridgelines when decaying through the $3/2^{-}$ state, while the solid lines indicate the ridgelines expected when decaying through the $1/2^{+}$ state.}
\label{fig:decay2plus2}
\end{figure}

Figure~\ref{fig:decay2plus2} compares the experimental 2$^+_2$ correlations to simulations of sequential decay via the $R$-matrix formalism \cite{lane1958} through the 1/2$^-$ [panel (c)] first-excited state and the 3/2$^-$ [panel (d)] fourth-excited state of $^{11}$N. Both simulations produced ridges in similar locations to the experimental data. In these simulations, one ridge is vertical, corresponding to the second emitted proton whose relative energy with the core is independent of angle, and a sloping ridge, corresponding to the first emitted proton whose relative energy to the core depends of the recoil imparted by the second proton and thus dependent on $\cos\theta_k$. In the experimental data, the differentiation of the ridges into a vertical and sloping varieties is not obvious.  It is possible that both $^{11}$N intermediate states are involved and wash out this differentiation. For example, the correlations plot in Fig.~\ref{fig:decay2plus2}(b) was obtained from incoherently adding the two decay paths with equal weight. If we have contribution from the 3/2$^-$ $^{11}$N intermediate
state, then as this state  has a decay branch to the excited state of $^{10}$C (Sec.~\ref{sec:N11}), this will contribute yield in the ``shoulder state'' (Sec.~\ref{sec:N11}).

\begin{figure}
\includegraphics[scale=0.4]{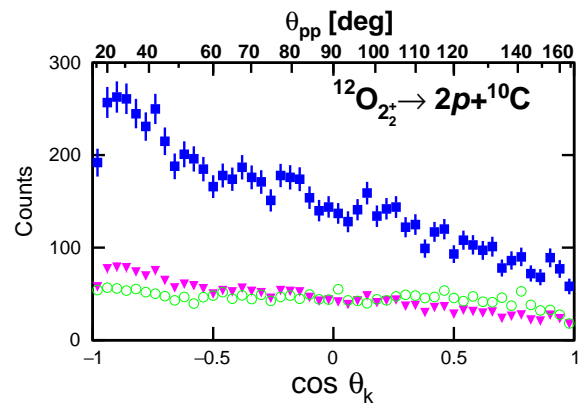}
\caption{Efficiency-corrected projection of the Jacobi Y correlation plot for the 2$^+_2$ $^{12}$O state on the $\cos\theta_k$ axis. Results are shown for (square data points) for a tight gate on the invariant-mass peak and for background gates on either side of this peak (triangles and open circles). An axis showing the relative angle $\theta_{pp}$ between the protons is also included.}
\label{fig:theta_k_tight}
\end{figure}

While there are sequential aspects of the 2$^+_2$ correlations in Fig.~\ref{fig:O12jacobis}(c), the yield along the ridges changes rapidly with $\cos\theta_k$ decreasing as $\cos\theta_k\rightarrow$1. If there is no memory between the two decay steps, one expects a distribution symmetric about $\cos\theta_k$=0, even if there is interference between the two proposed decay paths as the two intermediate states have the same parity. 
 Figure~\ref{fig:theta_k_tight} shows the projection of the Jacobi Y distribution, corrected for the detection efficiency, as the square data points. To reduce the background contribution, the data have been  selected with a  narrow gate (1.4-MeV wide) at the peak of the 2$^+_2$ state. To estimate the remaining background,  we placed a gate between the two 2$^+$ states (triangular data points) and one above the 2$^+_2$ state (open circular data points) where the fits suggest the background is dominant. These background distributions have been scaled to give the total background contribution as determined in our fits  and the two estimates are similar. The data show a strong $\cos\theta_k$ dependence with the maximum near where the relative angle between the protons is $\theta_{pp}$=0. The background distributions  cannot explain all the $\theta_{pp}$ dependence and thus the decay path deviates from the notion of independent sequential decays where the second decay has no memory of the first apart from the dictates of angular-momentum conservation. A similar observation was found for the 2$p$ decay of 2$^+_1$ state in $^{16}$Ne \cite{Brown:2015a} where the preference for small relative angles between the protons was explained  in the 3-body model as an interference between  sequential and prompt decay components. The sequential decay of the $^{17}$Ne second excited state also shows a preference for small $\theta_{pp}$ after the proton-proton final state interaction was considered, but the magnitude was much smaller than observed here \cite{Charity:2018a}. In the $^{17}$Ne case, the $^{16}$F intermediate state had a much smaller decay width ($\Gamma\sim$20~keV) than the two possible $^{11}$N intermediate states we determined for the decay of the 2$^+_2$ state. This suggests that this memory effect diminishes as the time between the two sequential decays increases.

\begin{figure}
\includegraphics[scale=0.4]{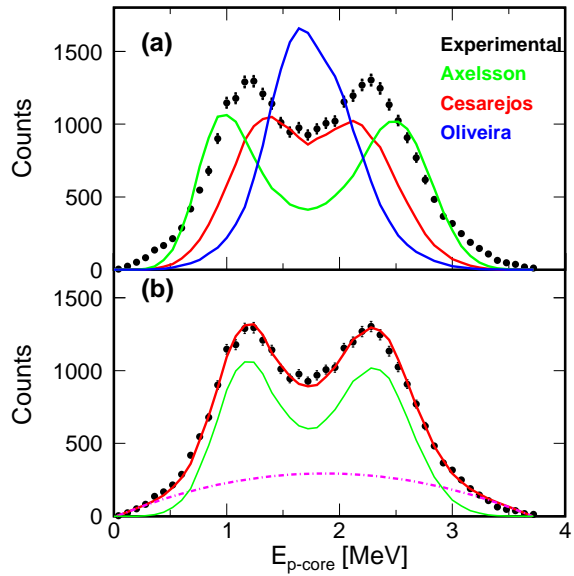}
\caption{Jacobi-Y energy projections of experimental (black dots) correlations
in the decay of  the $^{12}$O 2$^+_1$ state  with the gate $\cos\theta_k > 0.5$. The solid curves are simulations of the  sequential decay through $^{11}$N$_{g.s.}$. The three colored curves in (a) correspond to the different literature parameters given in Table~\ref{tbl:N11_gs}. The red solid curve in (b) corresponds to a fit to the present data.  The magenta dot-dashed curve corresponds to the fitted background, and the solid green curve corresponds to the total fit after subtraction of this background.}
\label{fig:2plus1Sequential}
\end{figure}

The correlations in Fig.~\ref{fig:O12jacobis}(b) for the 2$^+_1$ state are dominated by a banana-shaped structure in the diproton region ($\cos\theta_k<$ -0.5). A look at the upper half of the correlation plot reveals the presence of two clear ridges suggesting a sequential-like decay component. As mentioned before, the sloping ridge corresponds to the emission of the first proton. The sloping ridge is the one on the right in Fig.~\ref{fig:O12jacobis}(b), so this proton has more energy than the second one. As the decay scheme in Fig.~\ref{fig:level_2pB10} shows, the only level that the $2^{+}_{1}$ state can decay through 
to give the correct energy ordering is the $1/2^{+}$ ground state of $^{11}$N.
 These ridges are further investigated by the energy-projection of Fig.~\ref{fig:O12jacobis}(b) for $\cos\theta_k>$0.5 that is plotted as the data points in Fig.~\ref{fig:2plus1Sequential}. 

The resonance energy and width of $^{11}$N$_{g.s.}$ is not well-defined as seen by the discrepancy of level parameters from three studies \cite{axelsson1996,oliveira2000,casarejos2006} given in Table~\ref{tbl:N11_gs}. These values were used in the ENSDF evaluation \cite{ENSDF}. By examining the sequential-component, it might be possible to clear up some of the confusion about the decay parameters of this state. The curves plotted in Fig.~\ref{fig:2plus1Sequential} show sequential predictions  obtained with the different parameter sets from these three studies. The experimental data does not appear to agree with a decay through the $^{11}$N$_{g.s}$ as described by \cite{oliveira2000}. It is also difficult to claim agreement with either \cite{axelsson1996} or \cite{casarejos2006}, as the experimental peaks appear to lie directly between the two curves. Our data would suggest an intermediate resonance energy to those from \cite{axelsson1996,casarejos2006}.  We provide our own fit to the data in Fig.~\ref{fig:2plus1Sequential}(b) which gives a  decay energy of 1.378(5) MeV and a width of 0.78(1) MeV. A smooth background (dot-dashed curve) was added to enable the reproduction of the wings of the experimental distribution.

\begin{table}
\caption{Decay energies and widths reported in the literature and in this work for the proton-decaying 1/2$^{+}$ ground state of $^{11}$N. The evaluated ENDSF decay energy is a average of the first three values.}
\label{tbl:N11_gs}
\begin{ruledtabular}
\begin{tabular}{c c c c}
$E_{T}$ [MeV] & $\Gamma$ [MeV]  & Authors & Ref. \\
\hline
1.30(4) & 0.99$^{+100}_{-200}$ & Axelsson, \textit{et al.} & \cite{axelsson1996} \\
1.54(2) & 0.83(3) & Casarejos, \textit{et al.} & \cite{casarejos2006} \\
1.63(5) & 0.4(1) & Oliveira, \textit{et al.} & \cite{oliveira2000} \\
1.378(15)\footnotemark[1] & 0.78(1) & This work & \\
1.49(6) & 0.83(3) & ENSDF & \cite{ENSDF} \\
\end{tabular}
\end{ruledtabular}
\footnotetext[1]{Systematic uncertainty included in  this case to aid in comparison to the other measurements.}
\end{table}

\begin{table}
\caption{Structure of the first three 2$^+$ states predicted with the Gamow coupled-channel approach \cite{Wang:2019}}
\label{tbl:Wang}
\begin{ruledtabular}
\begin{tabular}{c c}
\multicolumn{2}{c}{2$^+_1$, $E_{T}$=3.803~MeV, $\Gamma$= 0.132~MeV}\\
65\% & ($s_{1/2},d_{5/2}$) \\
23\% & ($d_{5/2},d_{5.2}$) \\
11\% & ($s_{1/2},s_{1/2}$) \\
\hline
\multicolumn{2}{c}{2$^+_2$, $E_{T}$=5.150~MeV, $\Gamma$=1.027~MeV} \\

94\% & ($p_{1/2},p_{3/2}$) \\
1\%  & ($p_{1/2},p_{1/2}$) \\
\hline
\multicolumn{2}{c}{2$^+_3$, $E_T$=6.235~MeV, $\Gamma$=1.982~MeV} \\
55\% & ($s_{1/2},d_{3/2}$) \\
29\% & ($s_{1/2},d_{5/2}$) \\
4\%  & ($p_{3/2},p_{1/2}$) \\
2\%  & ($d_{5/2},d_{5/2}$) \\
2\%  & ($p_{3/2},p_{3/2}$) \\ 
2\%  & ($S_{1/2},s_{1/2}$) \\

\end{tabular}
\end{ruledtabular}
\end{table}

\subsection{Comparison to GCC predictions}

The structure of $^{12}$O states have recently been predicted with the Gamow coupled-channel approach \cite{Wang:2019} which treats $^{12}$O as two protons interacting with a deformed $^{10}$C core. The wavefunction of the protons is expanded using the Berggren basis, which includes bound and scattering states, thus considering the effect of the continuum. An initial comparison with the experimental levels of $^{12}$O was made in \cite{Wang:2019} and the predicted ground-state width was found to be similar to the experimental value. It is interesting to see if the sequential-decay components of the observed 2$^{+}$ states are consistent with the predicted internal structure. The wavefunction components predicted for the three 2$^+$ states are listed in  Table~\ref{tbl:Wang}. 

 The 2$^+_1$ state has a sequential component which decays through the $J^\pi$=1/2$^+$ ground state of $^{11}$N, consistent with the sequential removal of a $d$ and then an $s$ wave proton. This matches well with the largest component of the predicted wavefunction for this state. This component and the next largest component ($d_{5/2},d_{5/2}$) could also produce a sequential decay contribution through the 5/2$^+$ state in $^{11}$N (see Fig.~\ref{fig:level_2pB10}), but as the decay energy of the first proton would be small, the barrier penetration fraction appears to have  suppressed it.

Our second observed 2$^+$ state at $E_T$=6.5~MeV is similar in energy to both the GCC predictions for the second and third 2$^+$ state. The two proposed sequential components deduced from the correlation plot both involve the emission of two $p$-wave protons and thus suggest that our observed 2$^+_2$ level is also the second 2$^+$ state in the calculations, which is dominated by a ($p_{1/2},p_{3/2}$) structure. Indeed the two sequential decay paths can then be interpreted as differing in the order in which the $p_{1/2}$ and $p_{3/2}$ protons are emitted.

\section{CONCLUSIONS}

Levels in $^{11}$N, $^{12}$N, and $^{12}$O have been observed with the invariant-mass technique using single- and multiple-nucleon knockout reactions from a $^{13}$O beam at $E/A$=69.5~MeV. The charged-particle decay products produced from interactions with a $^9$Be target were detected  with the HiRA array.  The $J^\pi$=1/2$^-_1$ and 3/2$^-_1$ levels were observed in the $^{11}$N$\rightarrow p$+$^{10}$C invariant-mass spectra and the 3/2$^-$ state was deduced to have decay branches to the ground and first excited states in $^{10}$C. In addition a new state was observed in this channel at either $E^*$=1.073(10) or 4.416(10)~MeV.

  High-statistics spectra were obtained for the 2$p$ decay of $^{12}$O in which a second 2$^+$ state was observed for the first time. The 2$^+_1$ peak in this spectrum was found to have a low-energy shoulder which could have contributions from the 0$^+_2$, 1$^-_1$, or a second 2$p$ decay branch of the 2$^+_2$ state to the excited state in $^{10}$C. A higher-lying $^{12}$O state was observed in the 4$p$+2$\alpha$ exit channel which has contributions from a number of different decay paths including fission-like decays, i.e. $\alpha$+$^8$C$_{g.s.}$  and $^6$Be$_{g.s.}$+$^6$Be$_{g.s.}$.

  Analogs to these $^{12}$O states were also observed in $^{12}$N. The 2$p$+$^{10}$B invariant-mass spectrum was very similar to that for $^{12}$O$\rightarrow$2$p$+$^{10}$C with the same peak structure. These peaks were shown to arise from 2$p$ decay of $T$=2 states to the isobaric analog state in $^{10}$B.  Higher-lying $T$=2 states were also observed in the 2$p$+$\alpha$+$^6$Li spectrum corresponding to fission-like decay to $\alpha$+$^8$B$_{IAS}$ and $^6$Be$_{g.s.}$+$^6$Li$_{IAS}$.  An additional $^{12}$N peak was found at the same excitation in the 3$p$+$^9$Be channels that probably is also $T$=2, but results from the decay of a separate level. 

 The momentum correlations in the prompt 2$p$ decay of $^{12}$O$_{g.s.}$ were found to be almost identical to those measured previously in $^{16}$Ne$_{g.s.}$ consistent with prediction of the 3-body model of Grigorenko \textit{et al.} \cite{Grigorenko:2002}. These correlations were also found to be rather consistent with those for the 2$p$ decay of its isobaric analog state in $^{12}$N. The correlations for the 2$^+_2$ state of $^{12}$O show evidence of sequential decay through the 3/2$^{-}_1$ and/or the 1/2$^-_1$ intermediate states in $^{11}$N. However, the distribution of relative angles between the protons was strongly biased to small angles, qualitatively consistent with the results for the second excited states of $^{16}$Ne  \cite{Brown:2015a}. This may be a general feature of sequential decay through wide intermediate states.  

The correlations for the 2$^+_1$ state show the presence of a significant diproton-like component, but a sequential-decay component through the ground-state of $^{11}$N is also observed. The latter component allowed for a new determination of the $^{11}$N$_{g.s.}$. Decay through this intermediate state is consistent with the internal structure at the 2$^+_1$ level in $^{12}$O predicted with Gamow coupled-channels calculations. No deviation from the quadratic isobaric multiplet mass equation was found for the $A$=12 quintet.

\begin{acknowledgements}
This material is based upon work supported by the U.S. Department of Energy, Office of Science, Office of Nuclear Physics under award numbers DE-FG02-87ER-40316, DE-FG02-04ER-41320, DE-SC0014552, DE-SC0013365 (Michigan State University), DE-SC0018083 (NUCLEI SciDAC-4 collaboration), and DE-SC0009971 (CUSTIPEN: China-U.S. Theory Institute for Physics with Exotic Nuclei); and the National Science foundation under grant PHY-156556. J.M. was supported by a Department of Energy National Nuclear Security Administration Steward Science Graduate Fellowship under cooperative agreement number DE-NA0002135.     
\end{acknowledgements}


\end{document}